 \documentclass[aps,twocolumn]{revtex4}
 \usepackage{amsmath,amssymb}

 \usepackage{tabularx}

 \usepackage{epsfig}
 \usepackage{graphicx}

 \newcommand{\be}{\begin{equation}}
 \newcommand{\ee}{\end{equation}}
 \newcommand{\bea}{\begin{eqnarray}}
 \newcommand{\eea}{\end{eqnarray}}
 \newcommand{\nn}{\nonumber}

\begin{document}

\title{Vacuum energy in Einstein-Gauss-Bonnet anti$-$de Sitter gravity}

\author{Georgios Kofinas$^1$ and Rodrigo Olea$^2$}

\date{\today}

\address{~}

\address{$^{1}$Departament de F{\'\i}sica Fonamental,
Universitat de Barcelona\\ Diagonal 647, 08028 Barcelona, Spain}

\address{$^{2}$Centro Multidisciplinar
de Astrof\'{\i}sica - CENTRA, Departamento de F\'{\i}sica, Instituto
Superior T\'{e}cnico, Universidade T\'{e}cnica de Lisboa, Av.
Rovisco Pais 1, 1049-001 Lisboa, Portugal}

\begin{abstract}
A finite action principle for Einstein-Gauss-Bonnet anti$-$de
Sitter gravity is achieved supplementing the bulk Lagrangian by a
suitable boundary term, whose form substantially differs in odd
and even dimensions. For even dimensions, this term is given by
the boundary contribution in the Euler theorem with a coupling
constant fixed demanding the spacetime to have constant (negative)
curvature in the asymptotic region. For odd dimensions, the action
is stationary under a boundary condition on the variation of the
extrinsic curvature. A well-posed variational principle leads to
an appropriate definition of energy and other conserved quantities
using the Noether theorem, and to a correct description of black
hole thermodynamics. In particular, this procedure assigns a
nonzero energy to anti-de Sitter spacetime in all odd dimensions.

\end{abstract}

\maketitle
\vspace{-0.3cm} In recent years, several experiments carried out
suggest observational evidence for a positive value of the
cosmological constant \cite{lambda}. Nonetheless, from a
theoretical point of view, the idea of existence of extra
dimensions and alternative gravity theories does not rule out a
negative cosmological constant in a higher-dimensional spacetime.
On the contrary, for example, particular braneworld models induce
a zero or positive cosmological constant on a four-dimensional
Universe \cite{lisa}. \newline A negative cosmological constant is
also appealing because of the possibility of a profound connection
between anti-de Sitter gravity and a conformal field theory (CFT)
living on its boundary, that has attracted a considerable
attention in the literature \cite{maldacena}. Even though some
remarkable progress has been achieved on a rather case-by-case
basis, a general proof of this duality remains unknown. In that
context, the existence of a nonzero energy for anti-de Sitter
(AdS) vacuum spacetime in the gravity side may be helpful to
identify the corresponding CFT at the boundary. Indeed, in five
dimensions, the matching between the zero-point energy for
Schwarzschild-AdS black hole and the induced (Casimir) energy of a
precise boundary field theory is one of the best known examples
that realizes this bulk/boundary correspondence \cite{bala}.
General Relativity with AdS asymptotics requires a regularization
procedure in order to define a finite energy for the solutions of
the theory. Logically, a vacuum energy can appear only when this
mechanism does not invoke the substraction of a background
configuration as, e.g., in Hamiltonian formalism. That is the case
of the counterterms method \cite{henniskende, roberto}, where the
finiteness of the action and its energy-momentum tensor is
obtained by the addition of covariant functionals of the boundary
metric, constructed by solving the Einstein equations in a given
asymptotic form of the metric \cite{fg}. In spite this algorithm
provides the correct counterterms for many cases, in high enough
dimension it becomes rather cumbersome, what makes the full series
still unknown. It is clear that the inclusion of quadratic
curvature terms in the action will turn this method of
regularization even more complex. \vspace{-0.1cm}
\par
In this article, we introduce the boundary terms that make the
definition of energy in Einstein-Gauss-Bonnet gravity finite in
all dimensions. On the contrary to the standard approach, these
terms appear as given geometrical structures, where the
regularization procedure amounts to fix a single coupling constant
in a well-defined variational problem for precise asymptotic
conditions.
\par
In $D=d+1$ dimensions, the EGB action can be supplemented by a
boundary term $B_{d}$ \bea
&&\!\!\!\!\!\!\!\!\!\!\!\!\!\!\!I_{D}\!=\!\frac{1}{16\pi
G_{\!D}}\!\int_{\!M}\!\!d^{D}\!x
\sqrt{\!-\hat{g}}\,\Big[\!{\hat{R}}-2\Lambda+\alpha (\hat{R}_{\mu
\nu\kappa\lambda}\hat{R}^{\mu\nu\kappa\lambda}\!-\nn\\
&&\,\,\,\,\,\,\,\,\,\,\,\,\,\,\,\,\,\,\,\,\,\,\,\,\,\,\,\,\,\,\,\,\,\,\,\,\,\,\,\,\,\,\,\,\,\,\,\,
-4\hat{R}_{\mu\nu}\hat{R}^{\mu\nu}\!+\!\hat{R}^{2})\!\Big]+c_{d}\!\int_{\!\partial\!
M}\!\!\!\!B_{d}, \label{actio}\eea that regularizes either the
conserved quantities and the Euclidean action. Hatted quantities
stand for $D$-dimensional ones. It is well known that coupling a
Gauss-Bonnet term still gives second-order field equations, and in
an effective action of string theory it corresponds to the leading
order quantum corrections to standard gravity \cite{zwi}. Braneworld
cosmologies with curvature corrections have also been considered,
e.g. in \cite{gbcosmo, gbinduced}. For later convenience, we take
the above action in the language of differential forms
\begin{eqnarray}
I_{\!D} \!\!\!&=&\!\!\!\frac{1}{ 16\pi (\!D\!-\!2)!G_{\!D}}
\!\!\int_{\!M}\!\!\!\epsilon
_{\!A_{\!1}...A_{\!D}}\!\Big(\!\mathcal{\hat{R}}
^{\!A_{1}\!A_{2}}\!+\!\frac{D\!-\!2}{D\ell^{2}}e^{\!A_{1}}\!e^{\!A_{2}}\!\!\Big)
e^{\!A_{3}}\!...e^{\!A_{\!D}}  \nonumber \\
&&\!\!\!\!\!\!\!\!\!\!\!\!\!\!+\alpha (\!D\!-\!2)(\!D\!-\!3)
\epsilon _{A_{1}\!...\!A_{\!D}}
\mathcal{\hat{R}}^{\!A_{1}\!A_{2}}\mathcal{\hat{R}}^{\!A_{3}\!A_{4}}e^{\!A_{5}}\!...e^{\!A_{\!D}}\!+\!c
_{d}\!\int_{\partial M}\!\!\!\!\!B_{d}, \label{action}\end{eqnarray}
where the cosmological constant
$\Lambda\!=\!-\frac{(D-1)(D-2)}{2\ell^{2}}$, the orthonormal
vielbein $e^{A}\!=\!e^{A}_{\,\,\,\mu}dx^{\mu}$ and the curvature
2-form is defined as
$\mathcal{\hat{R}}^{AB}\!=\!\frac{1}{2}\hat{R}^{\mu\nu}_{\kappa\lambda}e^{A}_{\,\,\,\mu}e^{B
}_{\,\,\,\nu}dx^{\kappa}dx^{\lambda}$ in terms of the spacetime
Riemman tensor. Wedge products are omitted throughout. We will also
use the symbol $\kappa_{_{\!D}}\!=\![16\pi (D\!-\!2)!\,
G_{D}]^{-1}$. We will consider spaces of negative constant curvature
in the asymptotic region, which means that at the boundary
\begin{equation}
\hat{R}^{\mu\nu}_{\kappa\lambda}+\frac{1}{\ell_{\!e\!f\!\!f}^{2}}\delta^{[\mu\nu]}_{[\kappa\lambda]}=0,
\label{aads}\end{equation}  where the effective AdS radius is
$\frac{1}{\ell_{\!e\!f\!\!f}^{2}}\!=\!\frac{1\pm
\sqrt{1+8\Lambda\alpha^{\ast}}}{2(D-1)(D-2)\alpha^{\ast}}$ and
$\alpha^{\ast}\!=\!\alpha\frac{(D-3)(D-4)}{(D-1)(D-2)}$.
\par
In general, any spacetime metric can be put in Gauss-normal
coordinates
$ds^{2}=N^{2}(\rho)d\rho^{2}+h_{ij}(x,\rho)dx^{i}dx^{j}$. For AAdS
spaces in EGB gravity (\ref{aads}), the Fefferman-Graham expansion
of the metric is also valid and defined by
$N=\ell_{\!e\!f\!\!f}/2\rho$, $h_{ij}=g_{ij}(x,\rho)/\rho$ and the
expansion $g_{ij}\!=\!g_{(0)ij}\!+\!\rho g_{(1)ij}\!+\!\rho^{2}
g_{(2)ij}\!+\!...$, where $g_{(0)}$ represents the metric of the
conformal boundary $\rho=0$ \cite{FG-GB,STuniversal}. The extrinsic
curvature $K_{ij}=-\partial_{\rho} h_{ij}/2N$ takes the form
\vspace{-0.05cm} \bea &&
\!\!\!\!\!\!\!\!\!\!\!\!\!\!\!K^{i}_{j}\!=\!K_{j\ell}h^{\ell i}\!=\!
\frac{1}{\ell_{\!e\!f\!\!f}}\delta^{i}_{j}-\frac{\rho}{\ell_{\!e\!f\!\!f}}(g_{(0)}^{-1}
g_{(1)})^{i}_{j}+...,\label{kfg} \eea that means that at the
boundary  \be K^{i}_{j}=\frac{1}{\ell_{\!e\!f\!\!f}}\delta^{i}_{j}.
\label{umb}\ee In the variational problem of a gravity theory
$h_{ij}$ and $K_{ij}$ are independent variables. A well-posed action
principle is defined for suitable boundary conditions for these
fields that makes the on-shell action stationary. In the present
paper, for the odd-dimensional case, we will consider that at the
boundary the variations obey \be \delta K^{i}_{j}=0,
\label{mixed}\ee --what is compatible with fixing the conformal
metric on $\partial M$ \cite{miol}-- and introduce the appropriate
boundary term which makes the EGB action to reach an extremum.
Contrary to the standard Dirichlet approach to regularize the AdS
action, where the counterterms are constructed out only of intrinsic
quantities, the boundary terms proposed here explicitly contain also
the extrinsic curvature, making evident that we are not dealing with
the same action principle.
\par
In AdS gravity, odd and even dimensions have different essential
features (existence of Weyl anomalies, vacuum energy, etc.) that
suggest technical differences concerning their regularization (e.g.,
as in the standard holographic renormalization). As we shall see
below, the difference in the prescriptions to treat even and odd
dimensional cases is related to the existence of topological
invariants \cite{even,unpu}.
\par
\textbf{Odd-dimensional case.} Starting for conciseness with
$D\!=\!5$, the boundary term that leads to a well-defined variation
of the action is
\begin{eqnarray}
B_{4}\!\!\!&=&\!\!\!\epsilon _{\!A_{\!1}\!...\!A_{5}}\theta
^{\!A_{1}\!A_{2}}e^{\!A_{3}}\!\Big(\!
\mathcal{R}^{\!A_{4}\!A_{5}}\!\!+\!\frac{1}{2}\theta^{\!A_{4}}_{\,\,\,\,C}\theta^{C\!A_{5}}\!\!+\!\frac{1}{6
\ell_{\!e\!f\!\!f}^{2}}e^{\!A_{4}}e^{\!A_{5}}\!\Big)\nn\\
\!\!\!&=&\!\!\!\sqrt{\!-h}\,\delta _{[j_{1}\!j_{2}\!j_{3}]
}^{[i_{1}\!i_{2}i_{3}]}K_{i_{1}}^{j_{1}}\!\Big(\!
R_{i_{2}i_{3}}^{j_{2}j_{3}}\!-\!K_{i_{2}}^{j_{2}}\!K_{i_{3}}^{j_{3}}\!+\!\frac{
1}{3\ell_{\!e\!f\!\!f}^{2}}\delta _{i_{2}}^{j_{2}}\delta
_{i_{3}}^{j_{3}}\!\Big)d^{4}\!x,\nn
\end{eqnarray}
where the second fundamental form
$\theta^{AB}\!=\!n^{A}K^{B}-n^{B}K^{A}$, $K^{A}\!=\!K^{A}_{B}e^{B}$,
$K_{AB}\!=\!-h_{A}^{C}h_{B}^{D}n_{C;D}$, and $n^{A}$ is the unit
normal vector at the boundary. In the ordered frame
$(e^{1}\!=\!Nd\rho, e^{a}\!=\!e^{a}_{\,\,i}dx^{i})$ its only
non-vanishing components are
$\theta^{1a}\!=\!-K^{a}\!=\!-K^{i}_{j}e^{a}_{\,\,i}dx^{j}$.
$\mathcal{R}^{AB}$ and \textbf{$R^{ij}_{k\ell}$} are the curvature
2-form and the Riemann tensor of the boundary metric respectively.
An arbitrary variation of the action (\ref{action}) produces the
equations of motion plus a surface term
\begin{eqnarray}
&&\!\!\!\!\!\!\!\!\!\delta I_{5}
\!=\!\int_{\!M}\!(E.O.M.)\!+\!\int_{\!\partial\!
M}\!\!2\kappa_{_{5}} \epsilon _{abcd}\delta\!
K^{a}e^{b}(e^{c}e^{d}\!+\!12\alpha\hat{\mathcal{R}}^{cd}) \nonumber \\
&&\,\,\,\,\,+4c_{4}\epsilon _{abcd}\delta\!
K^{a}e^{b}\Big(\!\hat{\mathcal{R}}^{cd}\!+\!\frac{1}{3\ell_{\!e\!f\!\!f}^{2}}
e^{c}e^{d}\!\Big)\nn\\
&&\!\!\!\!\!\!\!-2c_{4}\epsilon _{abcd}( \delta\!
K^{a}e^{b}\!-\!K^{a}\delta e^{b}) \!\Big(\!
\mathcal{R}^{cd}\!-\!\frac{1}{2}K^{c}\!K^{d}\!+\!\frac{1}{2\ell_{\!e\!f\!\!f}^{2}}
e^{c}e^{d}\!\Big)\!.\label{var5}
\end{eqnarray}
The last term in (\ref{var5}) is canceled by the conditions
(\ref{umb}), (\ref{mixed}). Assuming the asymptotic behavior
(\ref{aads}) for the curvature, the coupling $c_{4}$ is fixed as
$c_{4}\!=\!(\ell_{\!e\!f\!\!f}^{2}\!-\!12\alpha)/128\pi G_{5}$ to
cancel the rest of the terms.
\par
For any odd dimension $D\!=\!2n+1$, the generalization of the
previous boundary term can be written in a compact form using
parametric integrations
\begin{eqnarray}
&&\!\!\!\!\!\!\!B_{2n}\!\!=\!n\!\!\!\int_{0}^{1}\!\!\!\!\!dt\!\!\int_{0}^{t}\!\!\!\!\!ds\,\epsilon
_{\!A_{\!1}\!...\!A_{\!2n\!+\!1}}\!\theta^{\!A_{\!1}\!\!A_{\!2}}\!e^{\!A_{3}}\!\Big(\!
\mathcal{R}^{\!A_{\!4}\!A_{\!5}}\!\!+\!t^{2}\theta^{\!A_{\!4}}_{\,\,\,C}
\theta^{C\!A_{\!5}}\!\!+\!\!\frac{s^{2}}{
\ell_{\!e\!f\!\!f}^{2}}\!e^{\!A_{\!4}}\!e^{\!A_{\!5}}\!\!\Big)\nn\\
&&\,\,\,\,\,...
\Big(\!\mathcal{R}^{\!A_{2n}\!A_{2n+1}}\!+\!t^{2}\theta
^{A_{2n}}_{\,\,\,\,\,\,F}\theta^{F\!A_{2n+1}}\!+\!\frac{s^{2}}
{\ell_{\!e\!f\!\!f}^{2}}e^{\!A_{2n}}e^{\!A_{2n\!+\!1}}\!\Big)\label{ena}\\
&&\!\!\!\!\!\!=\!2n\sqrt{\!-h}\!\!\int_{0}^{1}\!\!\!\!\!dt\!\!\int_{0}^{t}\!\!\!\!\!ds\,\delta
_{[\!j_{\!1}\!...\!j_{\!2\!n\!-\!1}\!]
}^{[i_{\!1}\!...\!i_{\!2\!n\!-\!1}\!]}K_{\!i_{\!1}}^{\!j_{\!1}}
\!\Big(\!{\frac{1}{2}}\!
R_{i_{2}i_{3}}^{j_{2}j_{3}}\!\!-\!t^{\!2}\!K_{i_{2}}^{\!j_{2}}\!K_{i_{3}}^{\!j_{3}}\!\!+\!\!\frac{
s^{2}}{\ell_{\!e\!f\!\!f}^{2}}\delta _{i_{2}}^{j_{2}}\!\delta
_{i_{3}}^{j_{3}}\!\!\Big)\nn \\
&&\!\!\!...\Big(\!{\frac{1}{2}}\!
R_{i_{2n\!-\!2}i_{2n\!-\!1}}^{j_{2n\!-\!2}j_{2n\!-\!1}}
\!\!-\!\!t^{2}\!K_{i_{2n\!-\!2}}^{j_{2n\!-\!2}}\!K_{i_{2n\!-\!1}}^{j_{2n\!-\!1}}\!\!+\!\!
\frac{s^{2}}{\ell_{\!e\!f\!\!f}^{2}}\delta
_{i_{2n-2}}^{j_{2n-2}}\delta
_{i_{2n-1}}^{j_{2n-1}}\!\!\Big)d^{2n}\!x.
\end{eqnarray} The on-shell variation of the action
(\ref{action}) is
\begin{eqnarray}
&&\!\!\!\!\!\!\delta\!I_{\!2\!n\!+\!1} \!\!=\!\!\!\int_{\!\partial\!
M}\!\!2\kappa_{_{\!D}} \epsilon
_{\!a_{\!1}\!...\!a_{\!2\!n}}\!\delta\!
K^{\!a_{\!1}}\!\Big(\!\!e^{\!a_{\!2}}e^{\!a_{\!3}}\!\!+\!\!2\alpha
(\!D\!-\!2\!)\!(\!D\!-\!3\!)\hat{\mathcal{R}}^{\!a_{\!2}a_{\!3}}\!\!\Big)
\!e^{\!a_{\!4}}\!\!...\!e^{\!a_{\!2\!n}}  \nn \\
&&\!\!\!\!\!\!\!+\!2nc_{\!2\!n}\!\!\!\int_{0}^{1}\!\!\!\!\!dt\epsilon
_{\!a_{\!1}\!...\!a_{\!2\!n}}\!\delta\!K^{\!a_{\!1}}\!e^{\!a_{\!2}}\!\Big(\!\!
\hat{\mathcal{R}}^{\!a_{\!3}a_{\!4}}\!\!+\!\!\frac{t^{2}}{\ell_{\!e\!f\!\!f}^{2}}\!
e^{\!a_{\!3}}\!e^{\!a_{\!4}}\!\!\Big)\!...\!\Big(\!\!
\hat{\mathcal{R}}^{\!a_{\!2\!n\!-\!1}\!a_{\!2\!n}}\!\!\!+\!\!\frac{t^{2}}{
\ell_{\!e\!f\!\!f}^{2}}\!e^{\!a_{\!2\!n\!-\!1}}\!e^{\!a_{\!2\!n}}\!\!\Big)  \nn \\
&&\!\!\!\!\!\!-\!2nc_{\!2\!n}\!\!\!\int_{0}^{1}\!\!\!\!dt\,
t\,\epsilon _{\!a_{\!1}\!...\!a_{\!2\!n}}\!(\!\delta\!
K^{\!a_{\!1}}\!e^{\!a_{\!2}}\!\!-\!\!K^{\!a_{\!1}}\!\delta\!
e^{\!a_{\!2}}\!)\!\Big(\!\!
\mathcal{R}^{\!a_{\!3}a_{\!4}}\!\!-\!t^{2}\!K^{\!a_{\!3}}\!K^{\!a_{\!4}}\!\!+\!\!\frac{t^{2}}
{\ell_{\!e\!f\!\!f}^{2}}\!e^{\!a_{\!3}}e^{\!a_{\!4}}\!\!\Big) \nn \\
&&...\Big(\!
\mathcal{R}^{a_{\!2n\!-\!1}a_{\!2n}}\!\!-\!t^{2}K^{\!a_{\!2n\!-\!1}}\!K^{\!a_{\!2n}}\!\!+\!\!\frac{t^{2}}{
\ell_{\!e\!f\!\!f}^{2}}e^{a_{\!2n\!-\!1}}e^{a_{\!2n}}\!\Big),
\end{eqnarray} and vanishes for the same boundary
conditions as in the five-dimensional case when \vspace{-0.2cm}  \be
c_{2n}\!=\!-\ell_{\!e\!f\!\!f}^{2n-2}\,\frac{\kappa_{_{\!D}}}{n}
\Big[\!1\!-\!\frac{2\alpha
}{\ell_{\!e\!f\!\!f}^{2}}(\!D\!-\!2\!)(\!D\!-\!3\!)\!\Big]
\!\Big[\!\int_{0}^{1}\!\!\!\!\!dt(t^{2}\!\!-\!1)^{n-\!1}\!\Big]^{\!-\!1}\!\!.
\!\!\vspace{-0.2cm}\ee {\bf Conserved quantities and vacuum energy.}
The charges $Q(\xi)$ associated to asymptotic Killing vectors $\xi$
--computed using the Noether theorem-- are defined as integrals on
the boundary of the spatial section $\Sigma$ at constant time. Their
expression is naturally split in two pieces \vspace{-0.2cm} \be
Q(\xi)=q(\xi)+q_{0}(\xi) \ee
\begin{eqnarray}
&&\!\!\!\!\!\!q(\xi)\!=\!-\!\!\int_{\!\partial\!\Sigma
}\!\!\!\kappa_{_{\!D}}\!\sqrt{\!-h}\,\epsilon
_{i_{\!1}\!...\!i_{\!2\!n}}\!(\!\xi^{k}\!K_{\!k}^{\!i_{\!1}}\!)\!\Big(\!\delta
_{[j_{\!2}\!j_{\!3}\!]\!}^{[i_{\!2}i_{\!3}\!]\!}\!+\!2\alpha
(\!D\!-\!2\!)(\!D\!-\!3\!)\hat{R}_{j_{\!2}j_{\!3}}^{i_{\!2}i_{\!3}}\!\Big)\nn\\
&&\,\,\,\,\,\,\,\,\,\,\,\,dx^{j_{\!2}}dx^{j_{\!3}}
dx^{i_{\!4}}\!...dx^{i_{\!2\!n}}
\!+\!2nc_{2\!n}\sqrt{\!-h}\!\!\int_{0}^{1}\!\!\!\!dt\,\epsilon
_{i_{\!1}\!...\!i_{2\!n}}\!(\!\xi^{k}\!K_{\!k}^{\!i_{\!1}}\!)
\delta_{\!j_{\!2}}^{i_{\!2}}\nn\\
&&\!\!\!\!\!\Big(\!\frac{1}{2}\!\hat{R}_{j_{\!3}j_{\!4}}^{i_{\!3}i_{\!4}}\!\!+\!\!\frac{
t^{2}}{\ell_{\!e\!f\!\!f}^{2}}\delta_{\!j_{\!3}}^{i_{\!3}}\!\delta
_{\!j_{\!4}}^{i_{\!4}}\!\Big)\!...\!\Big(\!
\frac{1}{2}\!\hat{R}_{j_{\!2\!n\!-\!1}j_{\!2\!n}}^{i_{\!2\!n\!-\!1}i_{\!2\!n}}\!\!+\!\!\frac{
t^{2}}{\ell_{\!e\!f\!\!f}^{2}}\delta
_{j_{\!2\!n\!-\!1}}^{i_{\!2\!n\!-\!1}}\delta
_{j_{\!2\!n}}^{i_{\!2\!n}}\!\Big)dx^{j_{\!2}}\!...dx^{j_{\!2\!n}}\\
 && \!\!\!\!\!\!\!q_{0}(\xi
)\!=\!2nc_{2n}\!\!\int_{\!\partial\!\Sigma
}\!\!\!\!\sqrt{\!-h}\!\int_{0}^{1}\!\!\!dt\,t\,\epsilon
_{i_{1}\!...\!i_{2n}}\xi^{k}(\delta_{j_{2}}^{i_{2}}K_{k}^{i_{1}}\!+\!\delta_{k}^{i_{2}}K_{j_{2}}^{i_{1}})\nn\\
&&\,\,\,\,\,\,\,\,\,\,\,\,\,\,\,\,\,\,\,\,\,\,\,\,\,\,\,\,\,\,\,\,\,\,\,\,\,\,\,\,\,\,\,\,\,\,\,\,
\Big(\!\frac{1}{2}\!
R_{j_{3}j_{4}}^{i_{3}i_{4}}\!-\!t^{2}K_{j_{3}}^{i_{3}}K_{j_{4}}^{i_{4}}\!+\!\frac{
t^{2}}{\ell_{\!e\!f\!\!f}^{2}}\delta _{j_{3}}^{i_{3}}\delta
_{j_{4}}^{i_{4}}\!\Big)...\nn\\
&&\Big(\!\frac{1}{2}\!
R_{j_{\!2\!n\!-\!1}j_{\!2\!n}}^{i_{2n\!-\!1}i_{\!2\!n}}\!-\!t^{2}\!K_{j_{\!2\!n\!-\!1}}^{i_{\!2\!n\!-\!1}}
\!K_{j_{\!2\!n}}^{i_{\!2\!n}}\!+\!
\frac{t^{2}}{\ell_{\!e\!f\!\!f}^{2}}\delta
_{j_{\!2\!n\!-\!1}}^{i_{\!2\!n\!-\!1}}\delta_{j_{\!2\!n}}^{i_{\!2\!n}}\!\Big)
dx^{j_{\!2}}\!...dx^{j_{\!2\!n}}\!.
\end{eqnarray}
It can be shown that $q(\xi)$ -in any odd dimension- can be
factorized by the l.h.s. of eq. (\ref{aads}), such that it
identically vanishes for spacetimes that satisfy (\ref{aads})
globally. Then, $q_{0}(\xi)$ provides a tensorial formula for the
vacuum energy for AAdS spacetimes.
\par
The static black hole solution of EGB gravity is \cite{bouldeser}
\vspace{-0.2cm}
\begin{equation}
ds^{2}=-\Delta^{2}(r)d\mathrm{t}^{2}+\frac{dr^{2}}{\Delta^{2}
(r)}+r^{2}\gamma _{\underline{m}\,\underline{n}}d\theta
^{\underline{m}}d\theta ^{\underline{n}}, \label{bd}\end{equation}
where
$\Delta^{2}(r)\!=\!k\!+\!\frac{r^{2}}{2(\!D\!-\!1\!)(\!D\!-\!2\!)\alpha^{\!\ast}}\!\!\Big[\!1\!\pm\!
\sqrt{1\!\!+\!8\Lambda\alpha^{\!\ast}\!\!+\!\!\frac{4(\!D\!-\!1\!)(\!D\!-\!2\!)
\alpha^{\!\ast}\!\mu}{r^{D-1}}}\Big]$, with $\mu$ appearing as an
integration constant, and $\gamma _{\underline{m}\,\underline{n}}$
($\underline{m},\underline{n}=1,...,D-2$) is the metric of the
transversal section $\Sigma_{\!D\!-\!2}^{k}$ of constant curvature
$k=\pm1, 0$. Solution (\ref{bd}) possesses an event horizon
$r_{+}$, which is the largest solution to $\Delta^{\!2}(r_{+})=0$.
For the time-like Killing vector $\xi=\partial_{\mathrm{t}}$, the
{\textit{mass}} is given by \vspace{-0.2cm}
\begin{eqnarray} &&\!\!\!\!\!\!
q(\partial_{\mathrm{t}})\!=\!M\!=\!vol(\Sigma_{\!D\!-\!2}^{k})
(D\!-\!2)!\,(\!\Delta^{\!2})^{\prime}\Big\{\!
\kappa_{_{\!D}}[r^{D-2}+\nn\\
&&\!\!\!\!\!\!\!+\!2\alpha
(\!D\!-\!2\!)(\!D\!-\!3\!)r^{\!D-4}(\!k\!-\!\Delta^{\!2}\!)]
\!+\!nc_{\!2\!n}r\!\!\int_{0}^{1}\!\!\!\!\!dt
\Big(\!\!k\!\!-\!\!\Delta^{\!2}\!\!+\!\!\frac{t^{2}\!r^{2}}{\ell_{\!e\!f\!\!f}^{2}}\!\Big)^{\!\!n\!-\!1}\!\Big\}
\!\Big|^{\!\infty}
\!\nn\label{m}\\
&&\,\,\,\,\,\,\,\,=\!\frac{(D\!-\!2)\,vol(\Sigma_{\!D\!-\!2}^{k})}{16\pi
G_{\!D}}\mu,
\end{eqnarray} (a prime stands for $d/dr$), while the {\textit{vacuum energy}}
is \vspace{-0.2cm}
\begin{eqnarray}&&\!\!\!\!\!\!q_{0}(\partial _{\mathrm{t}})\!=\!E_{0}
\!=\!2nc_{2n}(D\!-\!2)!\,vol(\Sigma_{\!D\!-\!2}^{k})\Big(\!
\Delta^{\!2}\!-\!\frac{r(\!\Delta^{\!2})^{\prime}}{2}\!\Big)\!\times\nn\\
&&\,\,\,\,\,\,\,\,\,\,\,\,\,\,\,\,\,\,\,\,\,\,\,\,\,\,\,\,\,\,\,\,\,\,\,\,
\,\,\,\,\,\,\,\,\,\,\,\,\,\,\,\,\,\,\,\,\, \int_{0}^{1}\!\!\!dt
\,t\,\Big(\!k\!-\!t^{2}\Delta^{\!2}\!+\!\frac{
t^{2}r^{2}}{\ell_{\!e\!f\!\!f}^{2}}\Big)^{\!n-1}\Big|^{\infty}\nn\label{e0}\\
&&\!\!\!\!\!\!\!\!=\!(\!-k\!)^{n}\frac{vol\!(\!\Sigma_{\!D\!-\!2}^{k}\!)}{8\pi
G_{\!D}}\ell_{\!e\!f\!\!f}^{2n\!-\!2}\frac{(\!2n\!-\!1\!)!!^{2}}{(2n)!}\!\Big(\!1\!-\!\frac{2\alpha
}{\ell_{\!e\!f\!\!f}^{2}}(\!D\!-\!2\!)(\!D\!-\!3\!)\!\Big)\!.
\label{vac}\end{eqnarray} The orientation $(e^{1},e^{0},
e^{\underline{m}})$ is used throughout. Expression (\ref{vac})
specialized for $D\!=\!5$ agrees with the vacuum energy obtained by
the surface counterterms method carried out in \cite{cvetic}. In the
limit of vanishing Gauss-Bonnet coupling, formula (\ref{vac})
reduces to the vacuum energy for Schwarzschild-AdS found with
Dirichlet counterterms \cite{roberto}.
\newline
{\bf Black Hole Thermodynamics.} The Euclidean period $\beta$ is
defined as the inverse of black hole temperature $T$ such that in
the Euclidean sector the solution (\ref{bd}) does not have a conical
singularity at the horizon and it is given by
$\beta=4\pi/(\Delta^{2})^{\prime}|_{r_{+}}$. In the canonical
ensemble, the Euclidean action $I^{\!E}\!=\!S\!-\!\beta \mathcal{E}$
defines the entropy $S$ and the thermodynamic energy
$\mathcal{E}\!=\!-\frac{\partial I^{E}}{\partial\beta}$ of a black
hole for a fixed surface gravity. The Euclidean bulk action is
evaluated for a static black hole of the form (\ref{bd})
as\vspace{-0.0cm} \bea
&&\!\!\!\!\!\!\!\!\!\!\!\!\!\!\!\!\!\!I^{E}_{bulk}\!=\!-\kappa_{_{\!D}}
(D\!-\!2)!\,vol(\Sigma_{\!D\!-\!2}^{k})\,\beta\,
\{\,(\!\Delta^{\!2})^{\prime}\,[r^{D-2}\nn\\
&&\,\,\,\,\,\,\,\,\,\,\,\,\,\,\,\,\,\,\,\,+2\alpha
(D\!-\!2)(D\!-\!3)r^{D-4}(k\!-\!\Delta^{\!2})]\,\}|_{r_{\!+}}^{\infty},
\label{dulce} \eea and the Euclidean boundary term as
\vspace{-0.2cm} \bea
&&\!\!\!\!\!\!\!\int_{\!\partial{\!M}}\!\!\!\!\!B_{2n}^{E}\!=\!-n(\!D\!-\!2\!)!\,
vol(\!\Sigma_{\!D\!-\!2}^{k}\!)\beta\Big[r(\!\Delta^{\!2})^{\prime}
\!\!\!\int_{0}^{1}\!\!\!\!dt\Big(\!k\!-\!\Delta^{\!2}\!+\!\frac{t^{2}r^{2}}
{\ell_{\!e\!f\!\!f}^{2}}\!\Big)^{\!\!n-\!1}\nn\\
&&\,\,\,\,\,\,\,\,\,\,+2\Big(\!\Delta^{\!2}\!-\!\frac{r(\!\Delta^{\!2})^{\prime}}{2}\!\Big)
\!\!\int_{0}^{1}\!\!\!dt\,t\,\Big(\!k\!-\!t^2\Delta^{\!2}\!+\!\frac{t^{2}r^{2}}
{\ell_{\!e\!f\!\!f}^{2}}\!\Big)^{\!n-\!1}\Big]\Big|^{\infty}\!.
\label{dul}\vspace{-0.2cm} \eea The contribution of the bulk action
$I^{E}_{bulk}$ at radial infinity plus the Euclidean boundary term
$c_{2n}\int_{\!\partial M}\!B^{E}_{2n}$ --by virtue of eqns.
(\ref{m},\ref{e0})--is equal to $-\beta(M\!+\!E_{0})$. The
finiteness of the charges ensures that the divergencies at
$r\!=\!\infty$ of the bulk Euclidean action are exactly canceled by
the ones coming from the boundary term. The thermodynamic energy
definition \vspace{-0.2cm}  \be \mathcal{E}=-\frac{\partial
I^{E}_{\!2n+1}/\partial r_{+}}{\partial \beta/\partial
r_{+}}=M+E_{0}, \label{thermoenergy}\vspace{-0.1cm}\ee  recovers the
same result for the total energy as from the Noether charges defined
above. Then, the entropy is the contribution of the Euclidean action
from the horizon \be
S=\frac{vol(\Sigma_{\!D\!-\!2}^{k})\,r_{+}^{D-2}}{4G_{\!D}}\Big[1\!+\!\frac{2k\alpha
(D\!-\!2)(D\!-\!3)}{r_{+}^{2}}\Big], \label{entro}\ee which agrees
with the result of \cite{cvetic} obtained by regularizing the action
with surface counterterms, and of \cite{ross} using the prescription
of \cite{iyer} at the horizon.
\par
\textbf{Even-dimensional case.} For $D=2n$ dimensions, the boundary
term is given by the (maximal) $n-$th Chern form \cite{chern}
\vspace{-0.1cm}
\begin{eqnarray}
&&\!\!\!\!\!\!\!\!B_{2n-1}\!=\!n\!\!\int_{0}^{1}\!\!\!\!dt\,\epsilon
_{A_{1}\!...\!A_{2n}}\theta^{\!A_{1}\!A_{2}}({\mathcal{
R}}^{\!A_{3}\!A_{4}}\!+\!t^{2}\theta _{\,\,\,\,C}^{\!A_{3}}\theta
^{C\!A_{4}})...\nn\\
&&\,\,\,\,\,\,\,\,\,\,\,\,\,\,\,\,\,\,\,\,\,\,\,\,\,\,\,\,\,\,\,\,\,\,\,\,\,\,\,
\,\,\,\,\,\,...({\mathcal{R}}^{\!A_{2n\!-\!1}\!A_{2n}}\!+\!t^{2}
\theta _{\,\,\,\,\,\,\,\,\,\,\,F}^{\!A_{2n\!-\!1}}\theta^{F\!A_{2n}}) \label{b1}\\
&&=\!2n\sqrt{\!-h}\!\int_{0}^{1}\!\!\!dt\,\delta _{[
i_{1}...i_{2n\!-\!1}]}^{[j_{1}...j_{2n\!-\!1}]}K_{j_{1}}^{i_{1}}\Big(
\frac{1}{2}
R_{j_{2}j_{3}}^{i_{2}i_{3}}\!-\!t^{2}K_{j_{2}}^{i_{2}}K_{j_{3}}^{i_{3}}\!\Big)...\nn\\
&&\,\,\,\,\,\,\,\,\,\,\,\,\,\,\,\,\,\,\,\, ...\Big( \frac{1}{2}
R_{j_{2n\!-\!2}j_{2n\!-\!1}}^{i_{2n\!-\!2}i_{2n\!-\!1}}
\!-\!t^{2}K_{j_{2n-2}}^{i_{2n-2}}K_{j_{2n-1}}^{i_{2n-1}}\!\Big)d^{2n\!-\!1}\!x\label{b3}.
\end{eqnarray}
The on-shell variation of the action (\ref{action}) is written as
\vspace{-0.1cm}
\begin{eqnarray}
&&\delta I_{2n}\!=\!\int_{\!\partial\!
M}\!\!2\kappa_{_{\!D}}\epsilon_{a_{1}...a_{2n-1}}\delta
K^{a_{1}}\Big(\!e^{a_{2}}e^{a_{3}}\!+\nn\\
&&\,\,\,\,\,\,\,\,\,\,\,\,\,\,\,\,\,\,\,\,\,\,+2\alpha
(D\!-\!2)(D\!-\!3)\hat{\mathcal{R}}^{a_{2}a_{3}}\!\Big)e^{a_{4}}...e^{a_{2n-1}}  \nonumber \\
&&+2n\,c_{2n-1}\,\epsilon _{a_{1}...a_{2n-1}}\delta
K^{a_{1}}\hat{\mathcal{R}}
^{a_{2}a_{3}}...\hat{\mathcal{R}}^{a_{2n-2}a_{2n-1}},
\end{eqnarray}
and vanishes for AAdS spacetimes (\ref{aads}) with an appropriate
choice of the coupling constant $c_{2n-1}$ \vspace{-0.1cm}
\begin{equation}
c_{2n-1}\!=(-1)^{n}\,\ell
_{\!e\!f\!\!f}^{2n\!-\!2}\,\frac{\kappa_{_{\!D}}}{n}\Big(\!1-\frac{2\alpha}{\ell
_{\!e\!f\!\!f}^{2}}(D\!-\!2)(D\!-\!3)\!\Big).
\vspace{-0.2cm}\end{equation} The asymptotic condition (\ref{aads})
ensures a well-defined action principle for even-dimensional EGB
gravity, that is the guiding line to achieve the finiteness of the
conserved charges constructed by the Noether theorem \vspace{-0.1cm}
\begin{eqnarray}
&&\!\!\!\!\!\!\!Q\!(\xi)\!=\!-\!\!\int_{\!\partial\!\Sigma
}\!\!\!\kappa_{_{\!D}}\!\sqrt{\!-h}\,\epsilon_{i_{\!1}\!...\!i_{\!2\!n\!-\!1}}\!(
\!\xi^{k}\!K_{\!k}^{\!i_{\!1}}\!)\!\Big(\!\!\delta
_{[\!j_{\!2}\!j_{\!3}]}^{[\!i_{\!2}i_{\!3}\!]}\!+\!2\alpha
(\!D\!-\!2\!)(\!D\!-\!3\!)\!\hat{R}_{j_{\!2}j_{\!3}}^{i_{\!2}i_{\!3}}\!\!\Big)\nn\\
&&\,\,\,\,\,\,\,\,\,\,\,\,\,\,\,\,\,\,\,\,\,\,\,\,\,\,\,\,\,\,\,\,\,\,\,\,\,
\,\,\,\,\,\,\,\,\,\,\,\,\,\,\,\,\,\,\,\,\,\,\,\,\,\,\,\,\,\,\,\,\,
\,\,\,\,\,\,\,\,\,\,\,dx^{j_{2}}dx^{j_{3}}dx^{i_{4}}\!...\!dx^{i_{2n-1}}\nn \\
&&\!\!\!\!\!\!\!+n\frac{c_{2\!n\!-\!1}}{2^{n\!-\!2}}\sqrt{\!-h}\,\epsilon
_{\!i_{\!1}...i_{\!2\!n\!-\!1}}(\!\xi^{k}\!K_{\!k}^{\!i_{\!1}}\!)
\hat{R}_{j_{\!2}j_{\!3}}^{i_{\!2}i_{\!3}}\!...
\hat{R}_{j_{\!2\!n\!-\!2}j_{\!2\!n\!-\!1}}^{i_{\!2\!n\!-\!2}i_{\!2\!n\!-\!1}}
\!dx^{j_{\!2}}\!...\!dx^{j_{\!2\!n\!-\!1}}\!.
\end{eqnarray}
The {\textit{mass}} for EGB-AdS black holes comes from the above
formula for $ \xi=\partial_{\mathrm{t}}$, that is \vspace{-0.1cm}
\begin{eqnarray}
&&\!\!\!\!\!Q(\partial _{\mathrm{t}})\!=\!M\!=\!vol(\Sigma
_{\!D\!-\!2}^{k})(D\!-\!2)!\,(\!\Delta^{\!2})^{\prime
}\Big\{\kappa_{_{\!D}} [r^{D-2}+\nn\\
&&+2\alpha(D\!-\!2)\!(D\!-\!3)r^{D-4}(k\!-\!\Delta^{\!2})]
\!+\!nc_{2\!n\!-\!1}(k\!-\!\Delta^{\!2})^{n-\!1}\!\Big\}\!\Big|^{\!\infty}\label{mm} \nn\\
&&\,\,\,\,\,\,\,\,\,\,\,=\!\frac{(D\!-\!2)vol(\Sigma_{\!D\!-\!2}^{k})}{16\pi
G_{\!D}}\mu, \vspace{-0.2cm}\end{eqnarray} expression that again
agrees with the standard results in the literature \cite{cai}. The
Euclidean continuation of the boundary term is now given by
\vspace{-0.1cm}
\begin{equation}
\int_{\!\partial\!
M}\!\!\!\!B_{2n\!-\!1}^{E}\!=\!-n(D\!-\!2)!\,vol(\Sigma
_{\!D\!-\!2}^{k})\,\beta\,(\!\Delta^{\!2})^{\prime}
(k\!-\!\Delta^{2})^{n\!-\!1}\Big|^{\infty}\!.  \!\label{dul}
\vspace{-0.2cm}\end{equation} The total Euclidean action
$I^{E}_{2n}=I_{bulk}^{E}+c_{2n-1}\!\int_{\!\partial\!M}\!B_{2n-1}^{E}$
at infinity corresponds to $-\beta M$, where $M$ is the Noether mass
computed above. The same expression for the mass is obtained from
the thermodynamic energy \vspace{-0.2cm}
\begin{equation} {\mathcal E}=-\frac{\partial I^{E}_{\!2n}/\partial
r_{+}}{\partial \beta/\partial r_{+}}=M\label{ftou}.
\end{equation} For the
entropy, the formula (\ref{entro}) is recovered for the $D\!=\!2n$
dimensional case \cite{ross}.
\newline
\textbf{Conclusions.} Unlike the usual methods to compute the
conserved quantities and entropy in EGB gravity, here we have
introduced for all dimensions the explicit form of the boundary
terms that remove the divergencies in the energy definition and
Euclidean action. We have also shown that in order to satisfy the
first law of black hole thermodynamics in odd dimensions, the energy
must appear as shifted by a vacuum energy respect to the mass
obtained in background-substraction methods. The general formula for
the energy of AdS vacuum might provide some insight on how gravity
induces a corresponding CFT on the boundary. \vspace{-0.1cm}
\par
In the standard regularization of AdS gravity, one can always add
a counterterm that is a local functional of the boundary metric.
It has been proposed \cite{ross} that this ambiguity might solve
the problem of negative values of eq.(\ref{entro}), shifting the
entropy by a constant. On the contrary, in the present approach,
additional boundary terms would in general spoil the AAdS boundary
conditions considered here. The only remaining freedom is the
possibility of adding --in even dimensions-- the Euler topological
term $\mathcal{E}_{2n}$ \cite{even} instead of the Chern form
(\ref{b1}). In that case, the entropy differs from the expression
(\ref{entro}) by a constant related to the Euler characteristic
$\chi(M)$ of the manifold, as noticed in \cite{OleaJHEP} for
EH-AdS gravity.

\vspace{-1.8cm}
\[ \]
{\bf Acknowlegements} The work of G.K. was supported by the European
Commission Marie-Curie Fellowship under contract
MEIF-CT-2004-501432. R.O. was supported by Funda\c{c}\~{a}o para a
Ci\^{e}ncia e Tecnologia (FCT) of the Ministry of Science, Portugal,
through project POCTI/FNU/44648/2002.

\end{document}